\documentclass[twocolumn,showpacs,nofootinbib,10pt]{revtex4-1}

\usepackage{color,float,amssymb,amsmath,upgreek}
\usepackage{color}
\usepackage{graphicx}
\usepackage[dvipsnames]{xcolor}
\usepackage{graphicx}\usepackage[colorlinks]{hyperref}
\newcommand{\approptoinn}[2]{\mathrel{\vcenter{
  \offinterlineskip\halign{\hfil$##$\cr
    #1\propto\cr\noalign{\kern2pt}#1\sim\cr\noalign{\kern-2pt}}}}}
\newcommand{\beq}{\begin{eqnarray}}
\newcommand{\eeq}{\end{eqnarray}}
\newcommand{\be}{\begin{equation}}
\newcommand{\ee}{\end{equation}}
\newcommand{\bea}{\begin{eqnarray}}
\newcommand{\eea}{\end{eqnarray}}

\newcommand{\ba}{\begin{eqnarray}}
\newcommand{\ea}{\end{eqnarray}}

\definecolor{green1}{RGB}{0,128,0} 
\hypersetup{hidelinks,backref=true,pagebackref=true,hyperindex=true,colorlinks=true,breaklinks=true,urlcolor= blue}
\hypersetup{%
  colorlinks = true,
  linkcolor  = blue,
  citecolor = green1,
}
\usepackage{bookmark,textgreek}
\usepackage{hyperref,color,xcolor}
\hypersetup{hidelinks,hyperindex=true,colorlinks=true,breaklinks=true,urlcolor= blue}
\hypersetup{%
  colorlinks = true,
  linkcolor  = blue
}

\begin{document}

\title{Dynamical tachyonic AdS/QCD and information entropy}

\author{N. Barbosa--Cendejas}
\email{nandinii@ifm.umich.mx} 
\affiliation{Facultad de Ingenier\'\i a
El\'ectrica, Universidad Michoacana de San Nicol\'as de Hidalgo.
Edificio $\Omega$, Ciudad Universitaria, C.P. 58040, Morelia,
Michoac\'{a}n, M\'{e}xico.}

\author{R. Cartas--Fuentevilla}
\email{rcartas@ifuap.buap.mx}
\author{A. Herrera--Aguilar}
\email{aherrera@ifuap.buap.mx}
\affiliation{Instituto de F\'{\i}sica, Benem\'erita Universidad
Aut\'onoma de Puebla. Apdo. Postal J-48, C.P. 72570, Puebla, M\'{e}xico}
\author{R. R. Mora--Luna}
\email{rigel@ifm.umich.mx}
\affiliation{Facultad de Ingenier'a Qu\'imica, 
Universidad Michoacana de San Nicol\'as de Hidalgo.
Edificio $M$, Ciudad Universitaria, C.P. 58040, Morelia,
Michoac\'{a}n, M\'{e}xico.}
\author{R.~da~Rocha}
\email{roldao.rocha@ufabc.edu.br}
\affiliation{Centro de Matem\'atica, Computa\c c\~ao e Cogni\c c\~ao, Universidade Federal do ABC - UFABC\\ 09210-580, Santo Andr\'e, Brazil.}

%
%


\begin{abstract}

The configurational entropy setup is employed to study  dynamical tachyonic holographic AdS/QCD models. The phenomenology of light-flavour mesonic states is then corroborated by the  Shannon's information. Tachyonic bulk corrections to the dynamical AdS/QCD 
show more dominant and abundant dual mesonic states in the 4D boundary QCD, when compared to the dual mesons in the standard dynamical AdS/QCD. Entropic Regge-like trajectories are also emulated.
\end{abstract}



\maketitle

\vspace*{-1cm}

\section{Introduction}

The quantum chromodynamics (QCD)  phenomenology, underlying  AdS/QCD holographic models, has been studied with a precise instrument called configurational entropy (CE) 
 \cite{Gleiser:2011di,Gleiser:2012tu}. Light-flavour mesons 
 \cite{Bernardini:2016hvx} and the glueball  phenomenology \cite{Bernardini:2016qit} were   investigated in the context of the CE, underlying the AdS/QCD setup. The CE was also used  to study and derive cross-sections of hadronic matter in the color-glass condensate in various contexts \cite{Karapetyan:2018oye,Karapetyan:2016fai,Karapetyan:2017edu}.
Besides, the quark-gluon condensate \cite{daSilva:2017jay} and quarkonia dissociation  \cite{Braga:2017fsb} were also explored in the context of the CE.

The CE  regards aspects of information in physical  systems \cite{Gleiser:2011di,Gleiser:2012tu}. Representing a certain logarithmic amount that devises information compression, the CE measures the entropy of shape of localized physical systems and configurations   \cite{Gleiser:2014ipa,Sowinski:2015cfa,Gleiser:2012tu}. 
More specifically, the CE contrives the compression of data inherent to modes in a given physical system, whose states, corresponding to lower values of the CE, shall  request less energy to be yielded, being also more abundant, more dominant, and also more probable to be detected and observed as well \cite{Bernardini:2016hvx}.  Further aspects of the CE were presented in Ref.  \cite{Gleiser:2018kbq}. 

A fruitful stage of applications for the CE is gravity. In fact, AdS$_5$-Schwarzschild black branes were studied in this context, corroborating to the Hawking--Page  phenomenon of phase transition between AdS$_5$-Schwarzschild black holes and thermal radiation, in Ref.  \cite{Braga:2016wzx}. Some features  of AdS charged black holes were studied in Ref. \cite{Lee:2017ero}. Besides, compact objects were scrutinized in Refs.   
 \cite{Gleiser:2013mga,Gleiser:2015rwa}, from the point of view of the CE. The (Chandrasekhar's) critical density, below which gravitons can condensate into compact systems, in the Bose--Einstein sense, was derived in Ref.   \cite{Casadio:2016aum} in the membrane paradigm. Moreover, elementary particle physics has been using the CE to derive further  phenomenological data \cite{Alves:2017ljt,Alves:2014ksa,Millan:2018fme}. The CE is also successful to study the thick brane paradigm and its physical aspects  \cite{roldao,Correa:2016pgr,Cruz:2017ems,Correa:2015lla}. A review of role of the Shannon's  informational entropy on heavy-ion collisions is presented in Refs. \cite{Ma:2018wtw,Ma:2014vsa,Ma:2015lpa}.

The AdS/QCD holographic setup presents an AdS$_5$  bulk, representing weakly coupled gravity, as the dual theory of the (strongly coupled) 4D conformal gauge field theory (CFT), on the boundary of AdS$_5$. Fields living in the AdS$_5$ bulk are dual to 4D QCD operators. In this context, the  confinement, preventing  color, charged,  particles to be  observed in an isolated way, can be then implemented in two ways, consisting either  of a hard cut-off along the extra dimension or of a dilatonic background that provides a dynamical (soft) cut-off \cite{Csaki}. Phenomenologically,  the quark-gluon plasma, the mesonic spectra and their decay constants, for example, were precisely predicted in the AdS/QCD correspondence, matching  to experimental data at the LHC and the RHIC  \cite{pdg1}.   
The QCD coupling constant increases with respect to energy decrements, implying that many hadronic features can not be  reported by a perturbative QCD. Hadronic phenomenology can be, thus, placed into the AdS/QCD correspondence, wherein the conformal symmetry may be broken by an energy parameter in the AdS bulk, which is dual to the infrared (IR) cut-off, governed by a mass scale, in the AdS$_5$ boundary gauge theory.  Besides the hard wall AdS/QCD setup \cite{Polchinski:2001tt,BoschiFilho:2002ta}, a 5D dilatonic bulk background was used in Ref. \cite{Karch:2006pv}, to implement the soft-wall setup,  that reproduces the Regge trajectories for mesons, according to the mass spectrum $m^2\underset{\sim}{\propto} S+n$, for excitation number $n$ and spin $S$. In the soft-wall,  a brane singularity is dynamically yielded by a scalar field emulating a smooth IR cut-off  \cite{Brodsky:2014yha}. 

AdS/QCD models were initially probed by the CE in Refs. \cite{Bernardini:2016hvx,Bernardini:2016qit,Braga:2017fsb}. Tachyonic 5D domain wall models can be implemented on smooth  5D branes \cite{Barbosa-Cendejas:2017vgm,German:2015cna,German:2012rv} within the framework of cosmological inflationary scenarios \cite{Barbosa-Cendejas:2017pbo, Barbosa-Cendejas:2015rba}.
 Now we want to explore the tachyonic corrections to dynamical AdS/QCD holographic models  \cite{Batell:2008zm}, from the point of view of their underlying informational content. 
We shall show how the tachyonic corrections to the  existing dynamical AdS/QCD models can derive a more compressed CE. Hence, the abundance of dual mesons shall be compared to their tachyonless counterparts.

In order to classify tachyonic corrections to the CE of excited light-flavour mesons, after briefly reviewing the CE setup,  in Sect. II the framework for the  dynamical tachyonic holographic AdS/QCD soft-wall is introduced.  Besides, their tachyonic counterparts have, respectively, a lower  CE for any fixed meson spin, corresponding to more abundant and dominant states. 
 The Sakai--Sugimoto model  and also higher spin mesons contributions, due to large $N_c$ suppression, is are used to compute the CE. Sect. III is devoted to provide our concluding remarks and outlook.

\section{configurational entropy of tachyonic AdS/QCD models}
\label{ILI}

The 5D action for the graviton-dilaton coupling in the Einstein frame reads
\begin{equation}\label{eframe}
\mathcal{S}=\int d^{5}x\sqrt{- g }\left[
\frac{1}{2}g^{\rm AB}\nabla_{\rm A} \upphi\nabla_{\rm B} \upphi+V(\upphi )-{R}\right] ,
\end{equation}
\noindent where we adopt hereon natural units; $R$ stands for the scalar curvature, and $%
\upphi$ denotes the dilaton field  depending on  the extra dimension, being $V(\upphi )$ the  
potential driving the model. A conformal coordinate system 
\begin{equation}
g_{AB}=e^{-2A(z)}\eta _{AB}, \label{metric}
\end{equation}
\noindent is employed, with $z = \int^y\exp\left(A(\d{y})\right)d\d{y}$ being the 5D energy scale, for $y$ denoting the fifth dimension, and $\eta_{AB}$ is the 5D Minkowski metric.  
The warp
factor 
\begin{equation}
A(z)=\log(z)+\mathring{A}\left( z\right) \label{dobra}
\end{equation} is necessary to match the AdS/QCD setup with experiments \cite{dePaula:2008fp}, for $\mathring{A}(z)$ running deformations
of the AdS$_{5}$ metric that are not conformal, where $\mathring{A}(0)=0$ yields the  asymptotic $\mathrm{AdS}_{5}$ bulk in the UV limit
 \cite{Karch:2006pv,dePaula:2008fp}. 

By denoting ${}^\prime$ the derivative with respect to the conformal $z$ coordinate, the Einstein--Euler--Lagrange equations for the action (\ref{eframe}) yield 
\begin{eqnarray}
3(A^{\prime \prime }(z)-A^{\prime }{}^{2}(z))-\frac{1}{2}\upphi ^{\prime
2}(z)-e^{-2A(z)}V(\upphi ) &=&0, 
\label{gremm1} \\
\upphi ^{\prime \prime }(z)-3A^{\prime }(z)\upphi ^{\prime }(z)-e^{-2A(z)}V_\upphi(\upphi)
&=&0, \label{fieldeqs}\\
6A^{\prime }{}^{2}(z)-\frac{1}{2}\upphi ^{\prime 2}(z)+e^{-2A(z)}V(\upphi ) &=&0, \label{gremm2}
\end{eqnarray}%
where $V_\phi(\phi)=\frac{dV}{d\upphi }$. The potential reads \cite{Kiritsis} 
\begin{equation}
V(\upphi \left( z\right) )=\frac32e^{2A\left( z\right)}\left[A^{\prime
\prime }\left( z\right) -3A^{\prime }{}^{2}\left( z\right) \right]\,.
\label{potentialconformal}
\end{equation}%

For studying the CE of light-flavour mesons, in the standard dynamical AdS/QCD setup, the  IR ($z\rightarrow \infty$) and UV ($z\rightarrow 0$) regimes of the metric in Eq. (\ref{metric}) must be appropriate ones. The warp factor  $\mathring{A}_{a }(z)\sim z^{a }$ in Refs. \cite{Kinar,Kruczenski}  is usually chosen, regarding the deformed warp factor in Eq. (\ref{dobra}). This choice 
is in full compliance for the dilaton to be a solution of (\ref{gremm1} -- \ref{gremm2}) and matches the one in Refs.  \cite{Kiritsis,dePaula:2010yu,dePaula:2009za,dePaula:2008fp} for $a =2$. Besides, Regge trajectories were studied in Ref. \cite{Karch:2006pv}, being here employed to investigate the CE of tachyonic AdS/QCD models and 
its phenomenological implementation in the context of 
light-flavour mesonic excitations. These excitations 
are emulated when Kaluza-Klein string modes	of  $S>2$ spin, related to the tensorial fields 
$\psi_{{\rm A}_1\ldots {\rm A}_S}$, are taken into account. The associated action $\int\! d^5x \!\sqrt{-g}\!\left[\nabla^A\psi^{{\rm A}_1\ldots {\rm A}_S}\nabla_{\rm A}\psi_{{\rm A}_1\ldots {\rm A}_S}\!\!+\!\!M^2\psi_{{\rm A}_1\ldots {\rm A}_S}\psi^{{\rm A}_1\ldots {\rm A}_S}\right]$ can derive the functions $\psi _{n,S}$, governed by the  
 Schr\"odinger-like equation,  
\begin{equation}\label{BBB1}\left[ -\partial _{z}^{2}+{\rm V}(z)\right] \psi_{n,S}=m_{n,S}^{2}\psi _{n,S},
\end{equation} with potential  
\begin{equation}\label{BBBB}
{\rm V}(z)=\frac14C^{{\prime}2}(z)-\frac12C^{\prime\prime }(z),\end{equation} with $C(z)=\left( 2S-1\right) A(z)+\upphi(z)$. The scalars $m_{n,S}^{2}$ correspond to the square of the 4D light-flavour  mesonic states mass  in QCD.  The Regge trajectories driven by Eq. (\ref{BBB1}) 
 have the mass spectra profile $m_{n}^{2}\sim n$ \cite{Karch:2006pv,Kiritsis}. For QCD data to be described \cite{pdg1}, the warp factor 
\begin{equation}
\mathring{A}(z,S)=f(S)\frac{z^2\Lambda^2}{%
e^{(1-z\Lambda)}+1}, \label{cosmologica}
\end{equation}%
where $f(S)=\frac{\sqrt{3}+1}{\sqrt{3}-1+2S}$ is then adopted, for $\Lambda\approx 300$ MeV \cite{dePaula:2010yu,dePaula:2008fp}. 

Tachyonic corrections to the dynamical AdS/QCD model can be then implemented, when the large $N_c$ limit of QCD represents the 4D holographic dual of the 5D  gravity,  coupled to both the 5D dilaton and the 5D tachyonic field. This system is described by a generalization of the action (\ref{eframe}), by including a tachyonic field
\begin{eqnarray}
S&=&\int d^5x \sqrt{-{g}} \Big[ \frac{1}{2}{g}^{{\rm AB}} \nabla_{\rm A}\upphi \nabla_{\rm B} \upphi\nonumber\\
&& +\frac{1}{2}{g}^{{\rm AB}}\nabla_{\rm A} T \nabla_{\rm B} T +{V}(\upphi,T)-{R} \Big],
\label{aef} 
\end{eqnarray} where $T$ denotes the tachyonic field,  associated with the closed string tachyon \cite{Batell:2008zm}. 
The dilaton and the tachyon are assumed to exclusively depend  on the fifth dimension \cite{De_Wolfe_PRD_2000,skend}. The following equations of motion   \cite{Batell:2008zm}, 
\begin{eqnarray}
12({A''}-2 {A'}^2) &=&{\upphi'}^2+{T'}^2+2{V}, \label{eom1}\\
24 {A'}^2&=&{\upphi'}^2+ {T'}^2-2 {V}, \label{eom2} \\
{\upphi''}-4 {A'}{\upphi'}& = &\frac{\partial {V}}{\partial \upphi},  \label{eom3}\qquad
{T''}-4 {A'}{T'} = \frac{\partial {V}}{\partial T}, \label{eom4} 
\end{eqnarray}
are, clearly, equivalent Eqs. (\ref{gremm1} -- \ref{gremm2}), in the absence of tachyonic fields.

Ref.  \cite{Batell:2008zm} shows that using conformal mappings for the dilaton, the metric have to 
be in the form  ${g}_{AB}=z^{-2}e^{-2 a z^2}\eta_{AB}$, where $a$ is an arbitrary constant, distinctly from Eq. (\ref{metric}). For it, new conformal coordinates are introduced, ${\color{black}{y(z)=\int^z dz'\, \frac{e^{-a z'^2}}{z'}}}=\frac{1}{2}{\rm Ei}\left(-a z^2\right),$ where ${\rm Ei}({\rm x})$ is the exponential integral function, yielding the warp function $
A(z)=a z^2+\frac{1}{4}\log({z}),$ and the dilaton and the tachyonic fields respectively read \cite{Batell:2008zm}
\begin{eqnarray}
\upphi(z)&=&\mp\,\sqrt{6} a z^2\,, \label{phi}\qquad 
T(z)=\pm \,6 \sqrt{a}z\,. \label{T}
\label{e}
\end{eqnarray}
The parameter $a$ has to be positive, to prevent a ghost  for the  tachyonic $T$ field.
The scalar potential reads \cite{Batell:2008zm}
\begin{eqnarray}
\!\!\!\!\!\!\!\!{V}^\pm(\upphi,T)&=&\frac{T^2}{2} \exp\left(\frac{T^2}{18}\right)+ 2\upphi^2 \exp\left(\mp\frac{\sqrt{6}\upphi}{3}\right)\nonumber\\
&&\!\!\!\!\!\!\!\!\!\!\!\!\!\!\!\!\!\!\!\!\!\!\!\!\!\!\!\!\!\!\!\!-2\sqrt{6} \left[3\sqrt{6} \exp\left(\frac{T^2}{36}\right)\!-\!2\left(\sqrt{6}\!\pm \!\upphi\right)\!\exp\!\left(\!\mp\frac{\sqrt{6}\upphi}{6}\right) \right]^2\!\!\!. 
\label{pot2}
\end{eqnarray}
 The choice $a=\sqrt{6}/6$ yields the AdS$_5$ metric, implementing the solution of Ref. \cite{Karch:2006pv}. Another possible solution $\upphi(z)\sim-z^2$ yields $g_{AB}=z^{-2}{\exp\left({-\frac{8 z^2}{3}}\right)}\eta_{AB}$ that is asymptotically AdS$_5$, in the UV regime.

The phenomenological scale $k$ can be introduced by  $z\mapsto  k z$, that can be identified, in the UV regime,  to the AdS$_5$ curvature. The parameter $a$ drives the location of the soft-wall $z_0=(\sqrt{a} k)^{-1}$, being the inverse of the mass scale. Besides, the potential (\ref{pot2}) can be expanded, up to second order in the fields,  as $
{V}^\pm(\upphi,T)\approxeq -(12+\frac{3T^2}{2}+2\upphi^2),$ being the first term related to the (negative) cosmological constant \cite{Batell:2008zm}. 
 Besides, Eq. (\ref{T}) yields all higher-order terms to be insignificant. 
 
Now we proceed to the computation of the CE for the above model. The CE relies on the  Shannon's  entropy in information theory.  Given a quadratically Lebesgue-integrable function
$\uprho(x)$ on the space $\mathbb{R}^D$ and its Fourier transform 
$\uprho(\omega) = \int_{\mathbb{R}^D}d^Dx\; \uprho(x)e^{-i\omega \cdot x},$ the modal fraction reads 
\cite{Gleiser:2012tu}
\begin{eqnarray}\label{modall}
\rho(\omega) = \frac{|\uprho(\omega)|^{2}}{\int_{\mathbb{R}^D} d^D\omega |\uprho(\omega)|^{2}},\label{modalf}
\end{eqnarray} comprising the weight of the system $\omega$ mode. The CE is then defined as  
\begin{eqnarray}
S[\rho] = - \int_{\mathbb{R}^D} d^D\omega\,{\rho}(\omega)\log {\rho}(\omega).\label{ce1}
\label{ce}
\end{eqnarray}
 The more information is needed for decoding any data, the less compressed  the information underlying the data shall be. In addition, the modal fraction in Eq. (\ref{modall}) regards  the shape of the configuration, in momentum space, to the  profile of $\uprho(x)$. Hence,  the CE represents the inherent informational content in general physical systems that have (localized) energy density configurations $\uprho(x)$. Hereon we adopt $D=1$, corresponding the $z$ energy scale in the AdS/QCD model. 

Taking into account the metric ansatz, the energy density reads $\rho=3e^{2A}(A''+2A'^2)\eta_{00}$ \cite{barb1,baz1}. 
Tachyonic corrections for the dynamical AdS/QCD holographic model can be then made compatible to  QCD data.  Eqs. (\ref{modalf}) calculated for this energy density are plot in Figs. 5 and 6 in the Appendix, for some values of the light-flavour meson spin, both in the IR and UV limits of Eq. (\ref{cosmologica}). Thus, Eqs. (\ref{modalf}) and \eqref{ce1} yield the CE of the dynamical tachyonic  AdS/QCD holographic model, for both the UV and IR limits, in  Fig. 1.
\begin{figure}[H]
\begin{center}
\includegraphics[width=2.7in]{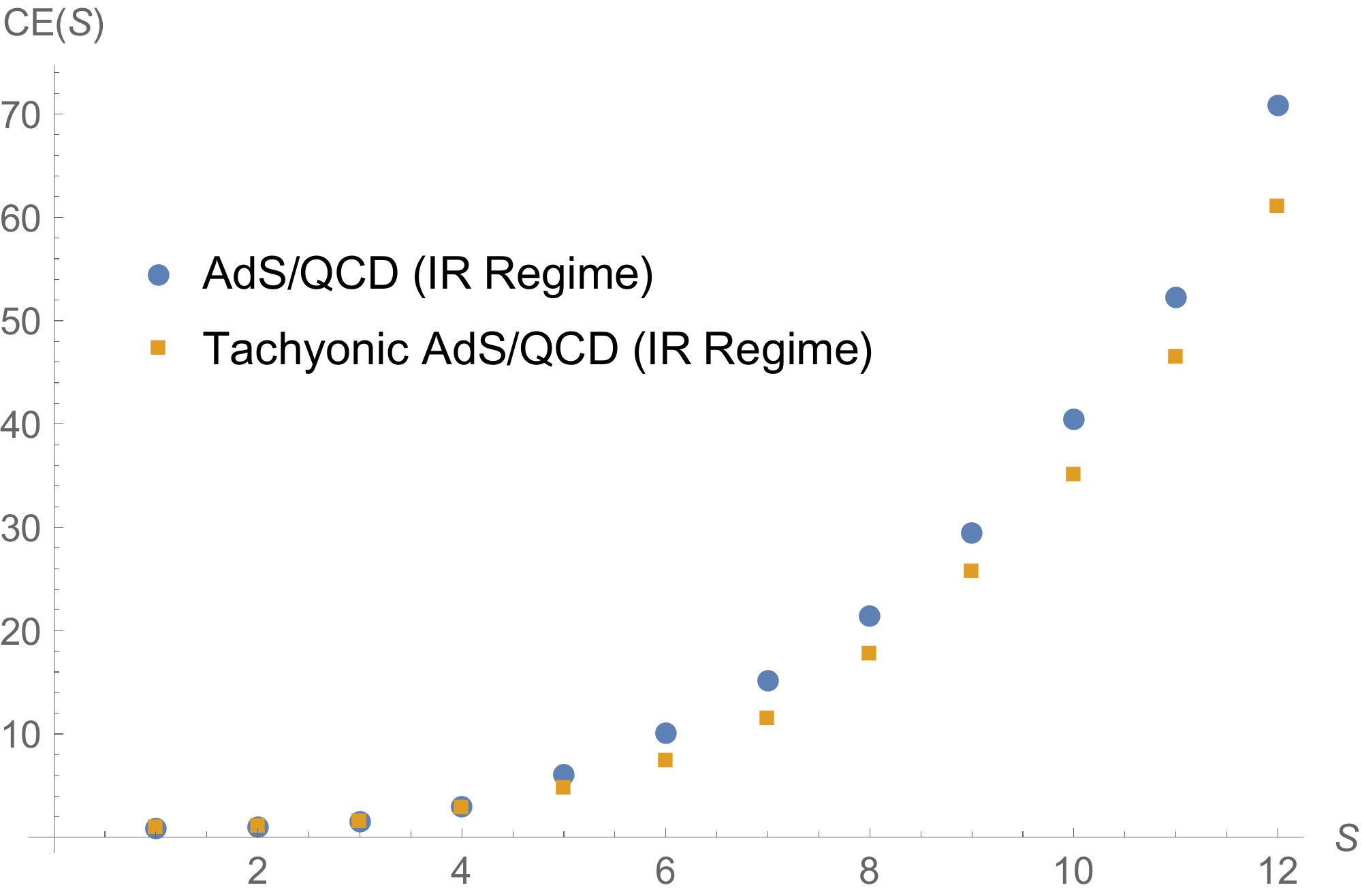}
\includegraphics[width=2.7in]{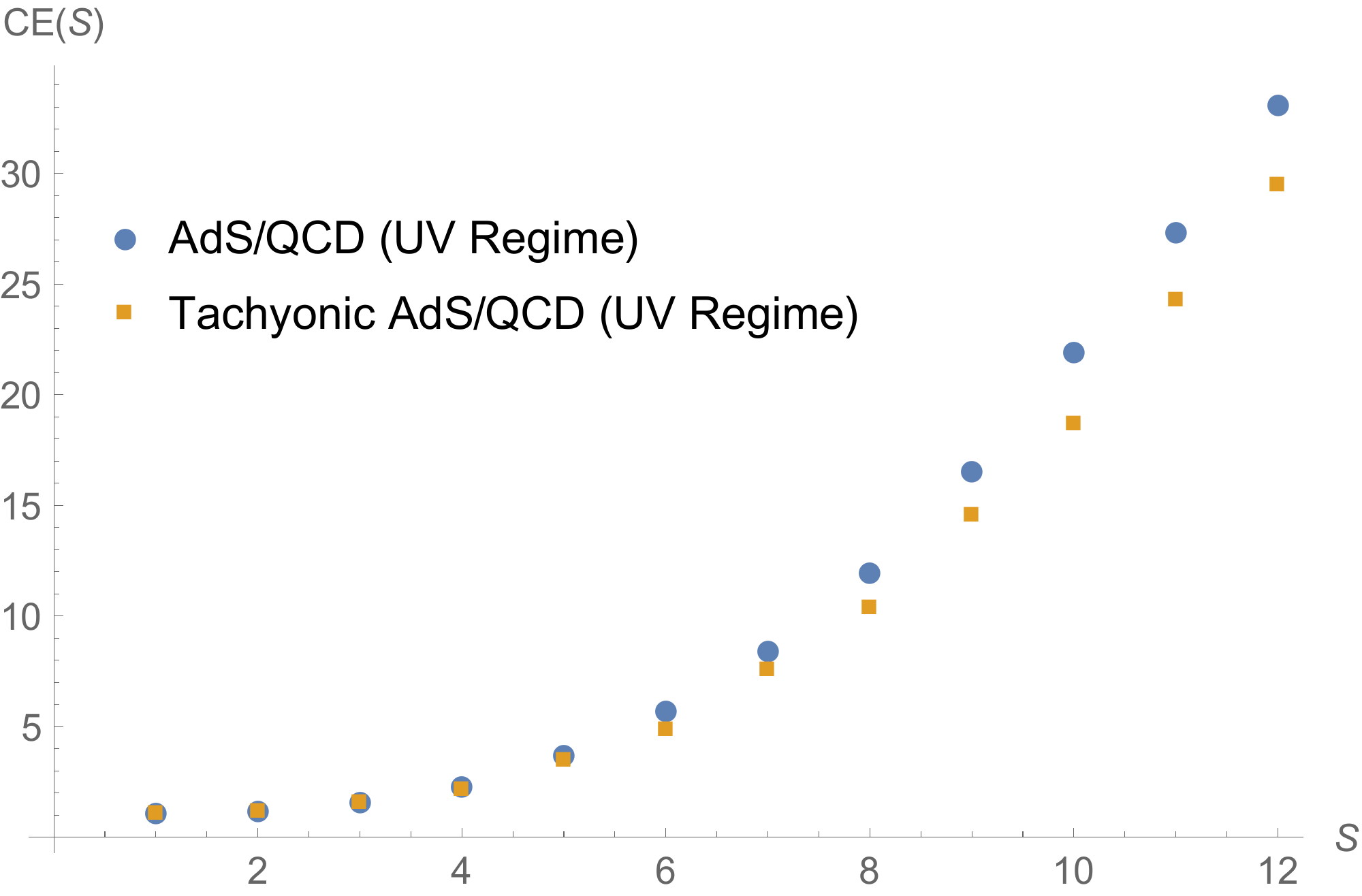}
\quad\quad
\caption{Configurational entropy (CE) $\times$ mesonic state spin $S$, for the IR regime, for the standard dynamical  AdS/QCD (blue dots) and its tachyonic corrections (yellow squares) [top panel];  for the UV regime, for the standard dynamical  AdS/QCD (blue dots) and its tachyonic corrections (yellow squares) [bottom panel].}
\end{center}
\end{figure}

The lower the light-flavour meson spin, the lower  the CE is, for both the UV and IR regimes. It reveals that the modes encrypted in the light-flavour mesons have a more compressed information underlying content, corroborating with the bigger abundance and dominance of lower spin mesonic states in Nature  that are  experimentally detected  \cite{Bernardini:2016hvx,Braga:2017fsb,pdg1}. 
Moreover, the fact that we have incorporated tachyonic corrections to the previous results in Ref. \cite{Bernardini:2016hvx} shows in Fig. 1 that the bulk tachyon field  (\ref{e}) 
   decreases the CE of the light-flavour mesons, for any fixed spin $S$. Still, as in the tachyonless case \cite{Bernardini:2016hvx}, the lower the mesons spin, the more predictable and dominant the informational  content of the system is. 
 Hence, light-flavour mesonic states are more dominant with tachyonic corrections, for both UV and IR regimes.

The value of CE can be still corrected 
for higher spin mesonic states, considering the  $D_4\!-\!D_8$ deformed soft-wall \cite{SongHe2010}. The (asymptotically) AdS$_5$ metric corresponds to the so called $D_3$ system, whereas the Sakai--Sugimoto
setup \cite{Sakai2005} regards the $D_4\!-\!D_8$ system \cite{BallonBayona:2010ae}, that can induce a deformation of the  soft-wall \cite{SongHe2010}, as $
A(z)\!=\!-{\rm a}_0 \log(z), \upphi(z)\!=\!{\rm d}_{0}\log(z)\!+\!{\rm c}_{0}z^{2}\,.$
The tachyon field in (\ref{e}) can be incorporated as well into the $D_4\!-\!D_8$ system by 
deforming $T(z)\!=\!\sqrt{6} z\!+\!f_0 \log(z)$.   
Besides, the function used to derive the light-flavour mesonic states in  Eq. (\ref{BBBB}) can be also deformed 
$
 C\!=\!\left( 2S\!-\!1\right) A\!+\!\upphi \mapsto C\!=\!-k (2S\!-\!1) A\!+\!\upphi,
$ 
for both the UV and IR \cite{SongHe2010}, where in the $D_4\!-\!D_8$ system one has $k\!=\!\frac73$, ${\rm a}_0\!=\!\frac32=-{\rm d}_0$, \cite{SongHe2010}, 
values adopted here to plot the graphics:
\begin{figure}[H]
\begin{center}
\includegraphics[width=2.8in]{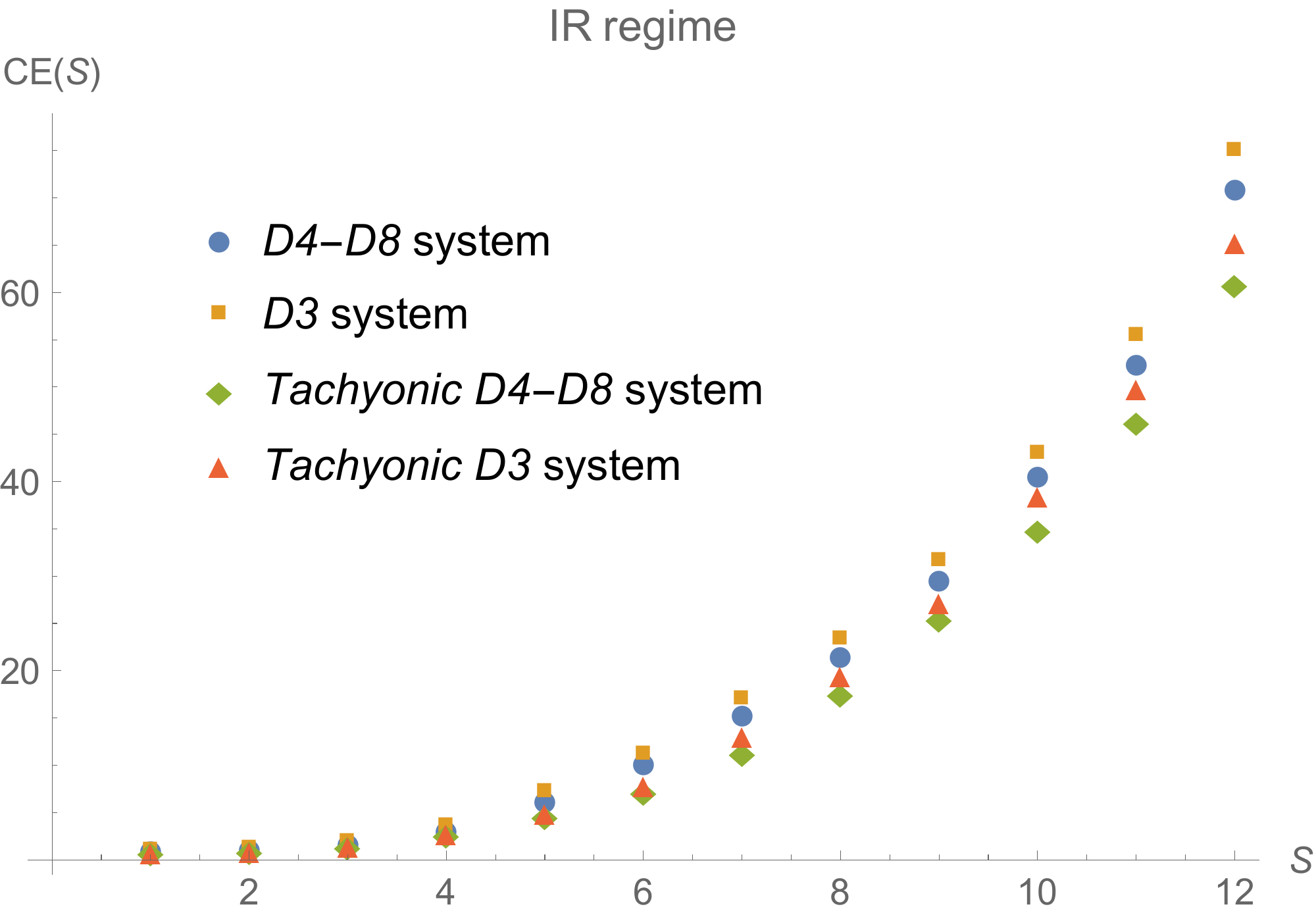}\\
\includegraphics[width=2.8in]{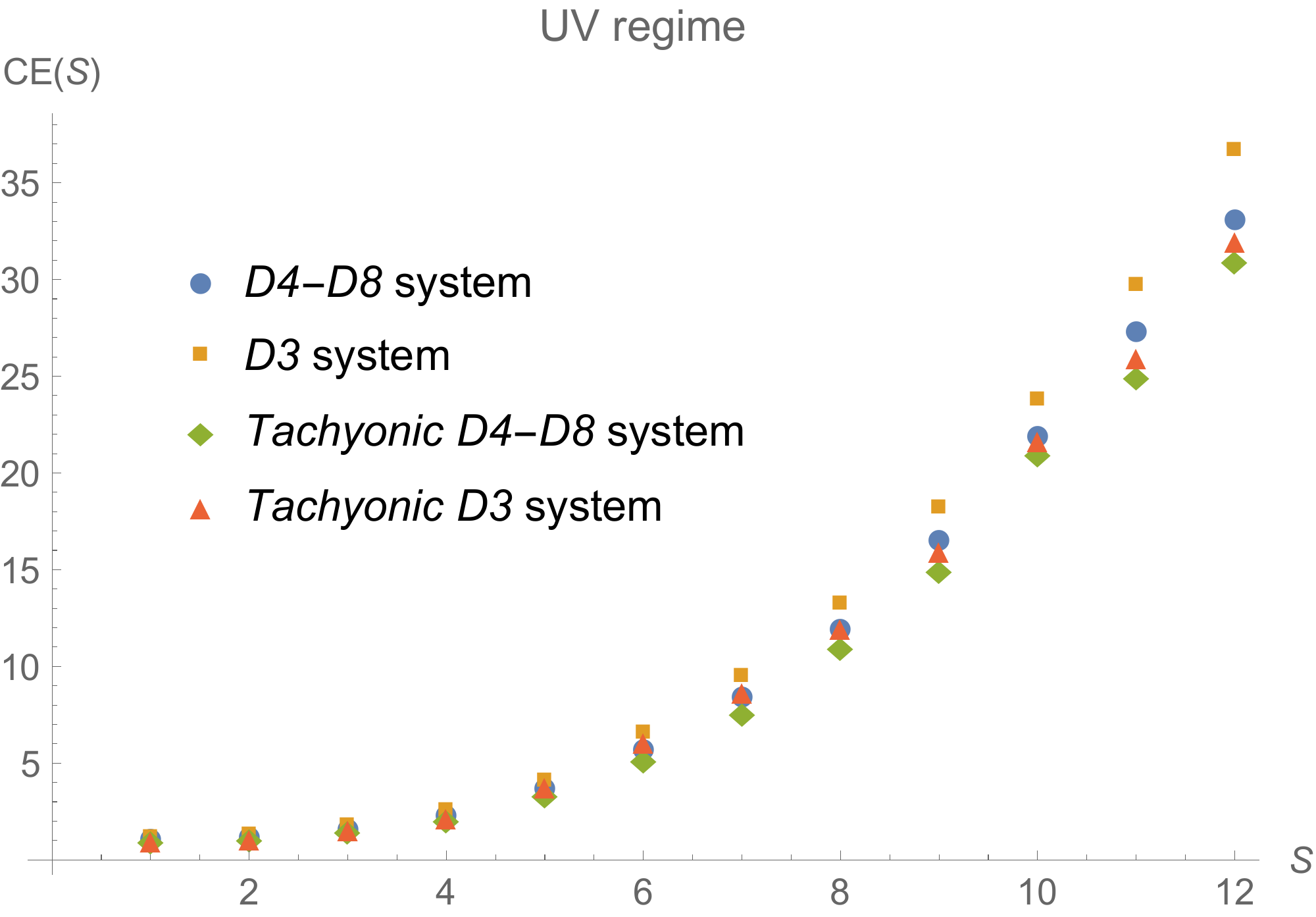}
\quad\quad
\caption{\footnotesize\; Configurational entropy (CE) $\times$ mesonic state spin $S$, of the standard AdS/QCD and its tachyonic corrections. The $D_3$ system (yellow squares) and its tachyonic corrections (red triangles) regard the AdS$_5$ metric, whereas the $D_4-D_8$ system (blue dots) and its tachyonic corrections (green diamonds) are also depicted for the IR [top panel] and UV [bottom panel] limits.}
\end{center}
\label{ce1001}
\end{figure}
Fig. 2, for the IR and UV regimes, respectively, evince the results of the CE of the light-flavour mesonic states, in the Sakai-Sugimoto model. 
It shows that the qualitative results are similar to the $D_3$ system. However, again tachyonic corrections imply a lower CE, compared with the tachyonless case, for each fixed spin. 
Besides, a quite interesting phenomenon occurs 
both in the $D_3$ and in the $D_4-D_8$ systems. In fact, let us observe the bottom panel of Fig. 2, corresponding to the UV asymptotic limit. For the $D_3$ system, mesonic states of spin $S=6$, with 
tachyonic corrections, have a lower value of the CE than the (tachyonless) mesonic states of spin $S=5$, suggesting that the tachyonic mesonic state $S=6$ may be more dominant than the (tachyonless) mesonic states of spin $S=5$. Their difference in the CE is just about $\sim0.5\%$.

Now, higher spin $S$ contributions, due to the large $N_c$ suppression \cite{nastase} can be also taken into account. Indeed, for higher values of the state spin  $S$, the energy density shall be corrected. The results are encoded in Fig. 3 below: 
\begin{figure}[H]
\begin{center}
\includegraphics[width=2.9in]{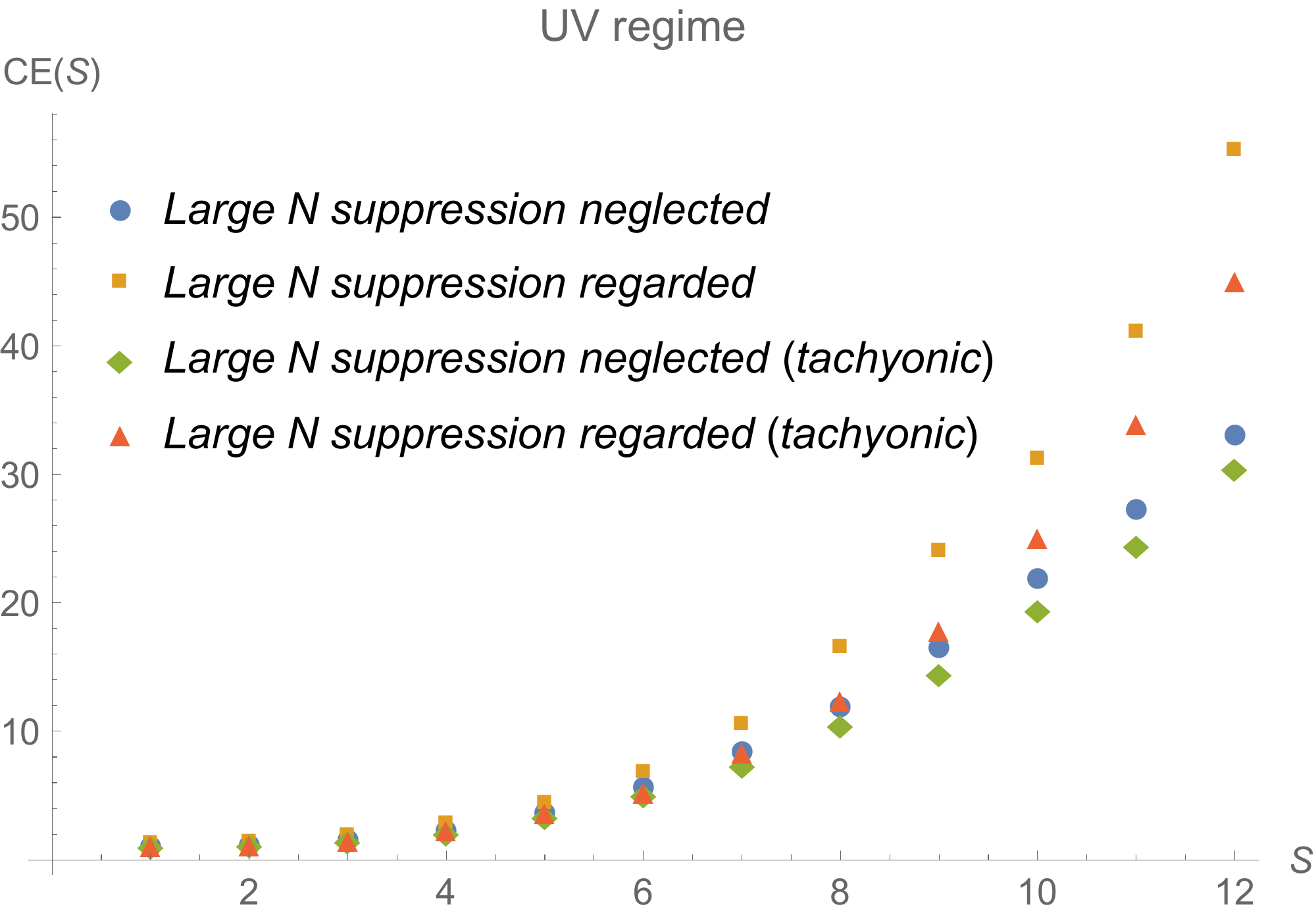}\\
\caption{\footnotesize\; Configurational entropy (CE) $\times$ mesonic state spin, neglecting and regarding the large $N_c$ suppression, for both tachyonic and tachyonless AdS/QCD model.}
\end{center}
\end{figure}
Again, in this case the higher the mesonic states spin, the higher the CE,  for both tachyonic and tachyonless AdS/QCD. However, the CE increases in a lower rate, for the tachyonic case.

\section{Concluding Remarks and Outlook}

Tachyonic potentials, as the one in the action (\ref{aef}), is motivated by a  $D$-brane -- anti-$D$-brane system, that presents 
a tachyonic excitation. At its minimum value, it contributes to the energy density and cancels out the sum of the tensions 
accumulated into the system \cite{Sen:1998sm}. Further aspects on the tachyon condensation can be seen, e. g., in Ref. \cite{Bonora:2011ri} in the context of the string field theory.

The  CE, underlying the dynamical tachyonic AdS/QCD model, was studied, showing that, also in the tachyonic background, light-flavour mesonic states with lower spins are more dominant and abundant than higher spin mesonic states. The most unexpected  result regards our analysis in Figs. 1, 2, and 3.
Starting from Fig. 1, for both the UV and the IR
regimes of AdS/QCD, the tachyon field (and its potential) in the 5D bulk action (\ref{aef}) for the 
tachyonic AdS/QCD model derives a CE for the dual mesons for any fixed spin, in the 4D boundary, that is lower than the tachyonless, standard, dynamical AdS/QCD. 
Besides, the UV limit in Fig. 1 has a lower CE than the IR limit, for any fixed mesonic state spin.
Fig. 2 brings also relevant results.
In fact, tachyonic corrections to 
the $D_3$ system in Fig. 2 provide a lower CE underlying the 4D mesonic states, when compared to their tachyonless counterparts. It occurs for both the IR and UV regimes. Besides, still in Fig. 2, the 
 Sakai--Sugimoto model, implementing a $D_4\!-\!D_8$ system, was studied, deforming the AdS/QCD soft-wall model. Again, for any fixed spin, the 4D light-flavour mesons present a lower CE, with  tachyonic corrections, when compared to the mesons without bulk tachyon field. In Fig. 3, 
the large $N_c$ supression was also considered.
Once more the tachyonic corrections to the AdS/QCD makes a lower CE of the mesonic states, for any fixed spin, when compared to the absence of bulk tachyon fields. In all the considered cases, adding a tachyonic field in the 5D bulk induces 
 more dominant and abundant light-flavour dual mesons in the 4D boundary. Besides, although the parameter $a$ is indeed a free parameter, the dynamical AdS/QCD model can be solely recovered by the choice $a=\sqrt{6}/6$, in Eqs. (\ref{T}). In fact, the factor $\sqrt{a}$ in the tachyon field $T$ in Eq. (\ref{T}) implies that the $a>0$, to prevent ghosts involving the $T$ field. Besides, the choice $a=\sqrt{6}/6$ implies that  the bulk metric corresponds to a pure AdS$_5$ space, being the dilaton quadratic in $z$, thus implementing the dynamical AdS/QCD model. Other values for the $a$ parameter  
might correspond to asymptotic AdS$_5$ spaces. Since the soft-wall location $z_0=(\sqrt{a} k)^{-1}$ depends on the $a$ parameter   and on the AdS$_5$ 
curvature $k$ as well, for the bulk cosmological constant of order $M^3k^2$,  where $M$ is the 5D Planck scale, one takes $1/z_0$ as an energy
scale. Experimental data can, thus, drive the range for the $a$ parameter \cite{pdg1}, for the mesonic mass spectra lying in the range 134.9 MeV/$c^2 \lesssim m\lesssim  11.01 {\rm GeV}/c^2$. Values of $a$ other than $\sqrt{6}/6$ might lead to other dynamical tachyonic models, that are not asymptotically AdS.

 Taking into account the mesons Regge trajectories \cite{Karch:2006pv} and experimental data  \cite{pdg1}, there is a correlation between the light-flavour mesonic states spin excitations and their CE. The higher the spin, the higher the CE is. Mesonic states with very high spins are predicted to be too rare to be produced and detected. In fact, their entropy of shape is extremely high, for very high values of the spin.  Hence, the CE is a valuable 
 instrument, both  for theoretical and experimental physics, to probe QCD phenomenology. 
 As the experimental data of  high-spin mesonic states is not prolific \cite{pdg1}, the CE can puzzle out  the categorization of high-spin light-flavour mesonic states into families. 
We conclude that the organizational information,  underlying the 4D boundary mesons, is even more compressed when a tachyonic field is included in the AdS/QCD action. Hence, tachyonic corrections to the CE  of light-flavour mesonic states are more dominant and abundant than their tachyonless counterparts,  for each fixed spin. 
 
 {\color{black}{Finally,  entropic Regge-like trajectories can be emulated from the dynamical AdS/QCD model, relating the CE to the mesonic state spin. It is worth to emphasize that the CE can be interpolated by the cubic function for the IR  regime in  tachyonless case, by  
 CE($S$) = $0.0451 \,S^3-0.1084\, S^2+0.6413\, S-0.0821$, at the given range of mesonic states spin, within 0.3\%. Besides, the logarithm of the CE  can be also interpolated by the equation  CE($S$) = $0.0385\, S^3-0.0427\, S^2-0.0372 \,S+0.8721$, for the tachyonic IR regime, within 0.2\%.  Analogously, the tachyonless UV regime case gives 
CE($S$) = $ 0.0025 \,S^3+0.2871 \,S^2-1.2122 \,S+2.2879$, within $0.3\%$, whereas 
 the tachyonic UV regime reads the relation 
 CE($S$) = $ 0.0365 \,S^3+0.0363 \,S^2-0.3420 \,S+1.1660$, within $0.3\%$,  
 relating the CE to the meson spin. \vspace*{0.5cm}
 \begin{figure}[H]
\begin{center}
\includegraphics[width=2.8in]{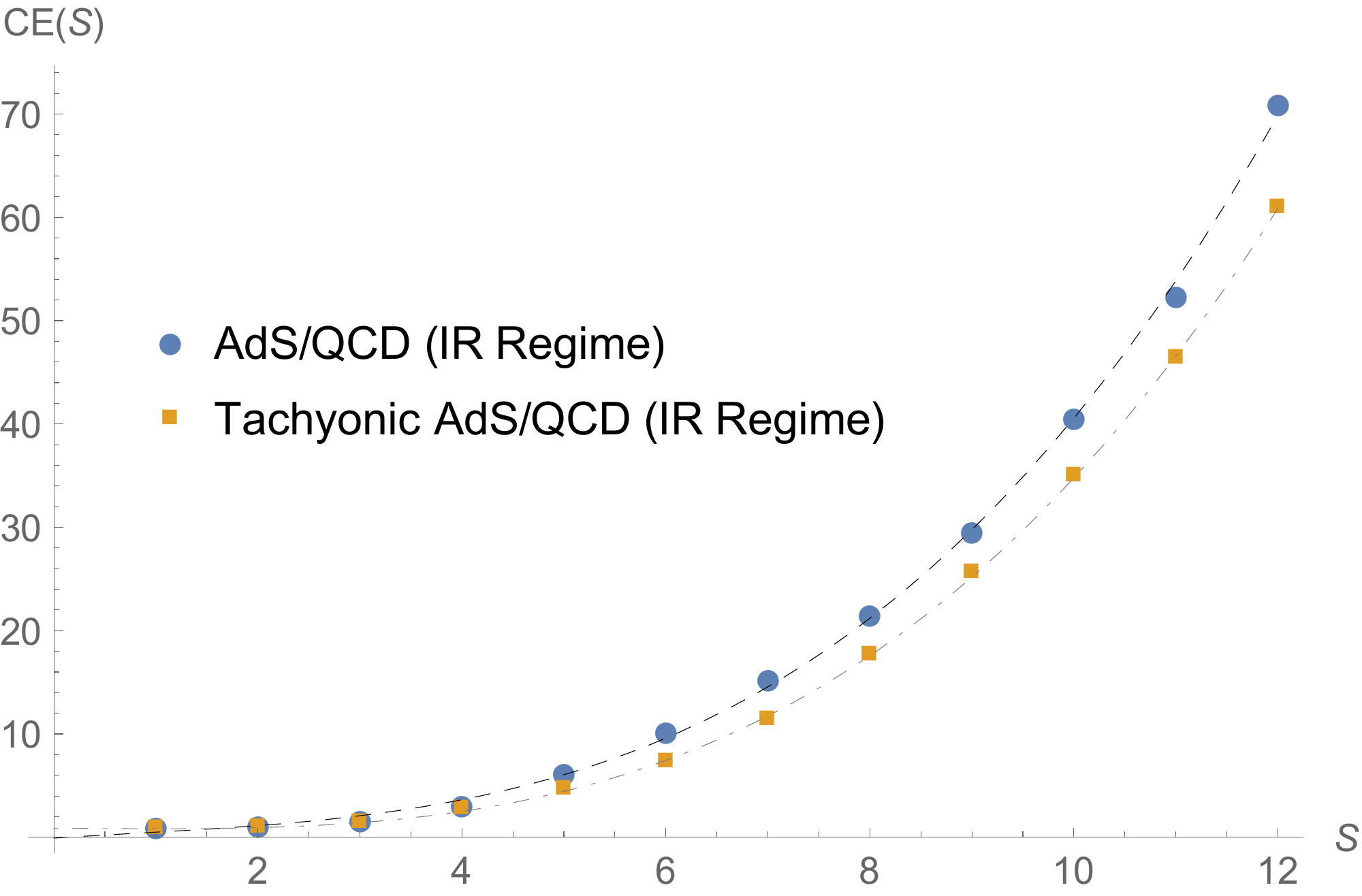}
\includegraphics[width=2.8in]{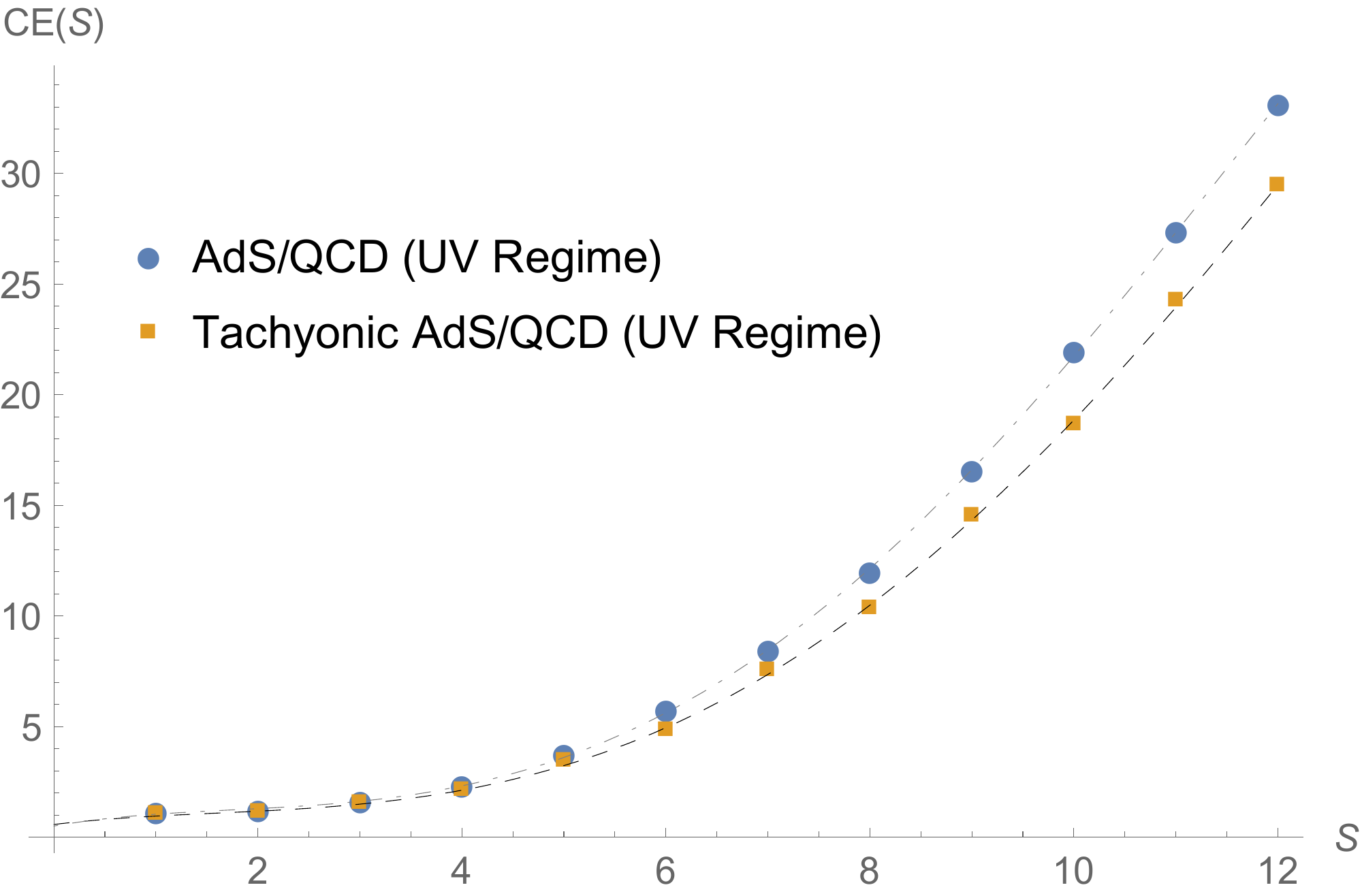}
\quad\quad
\caption{Interpolation of the CE as a function of the light-flavour meson spin $S$, for the IR and UV regimes.}
\end{center}
\end{figure}}}
 As AdS/QCD is an effective low energy limit of AdS/CFT, 
 a further hard task may consist of investigating supergravity effective domain wall solutions in the context of the CE \cite{Cvetic:2002su}.
\acknowledgments
 
NBC thanks to FIE--UMSNH;   AHA and RCF are grateful to a VIEP-BUAP grant; NBC, RCF, AHA, and RRML thank SNI for financial support. RdR~is grateful to to FAPESP (Grant No.  2017/18897-8) and to CNPq (Grant No. 303293/2015-2), for partial financial support.
\vspace*{-2cm}
	\begin{widetext}\appendix 
	\section{Some modal fractions used in the CE computations}
\begin{figure}[H]
\begin{center}
\includegraphics[width=2.5in]{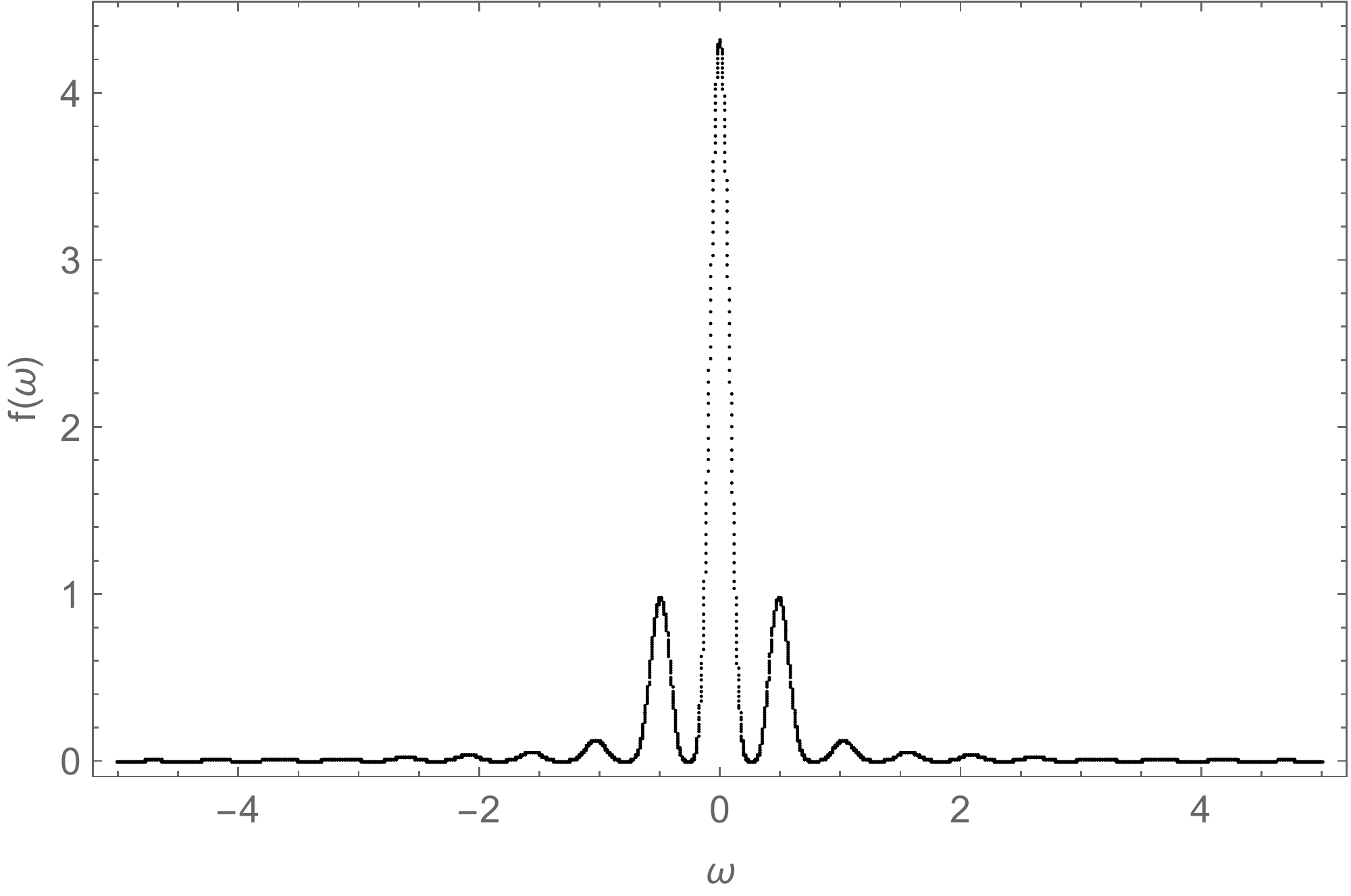}
\includegraphics[width=2.6in]{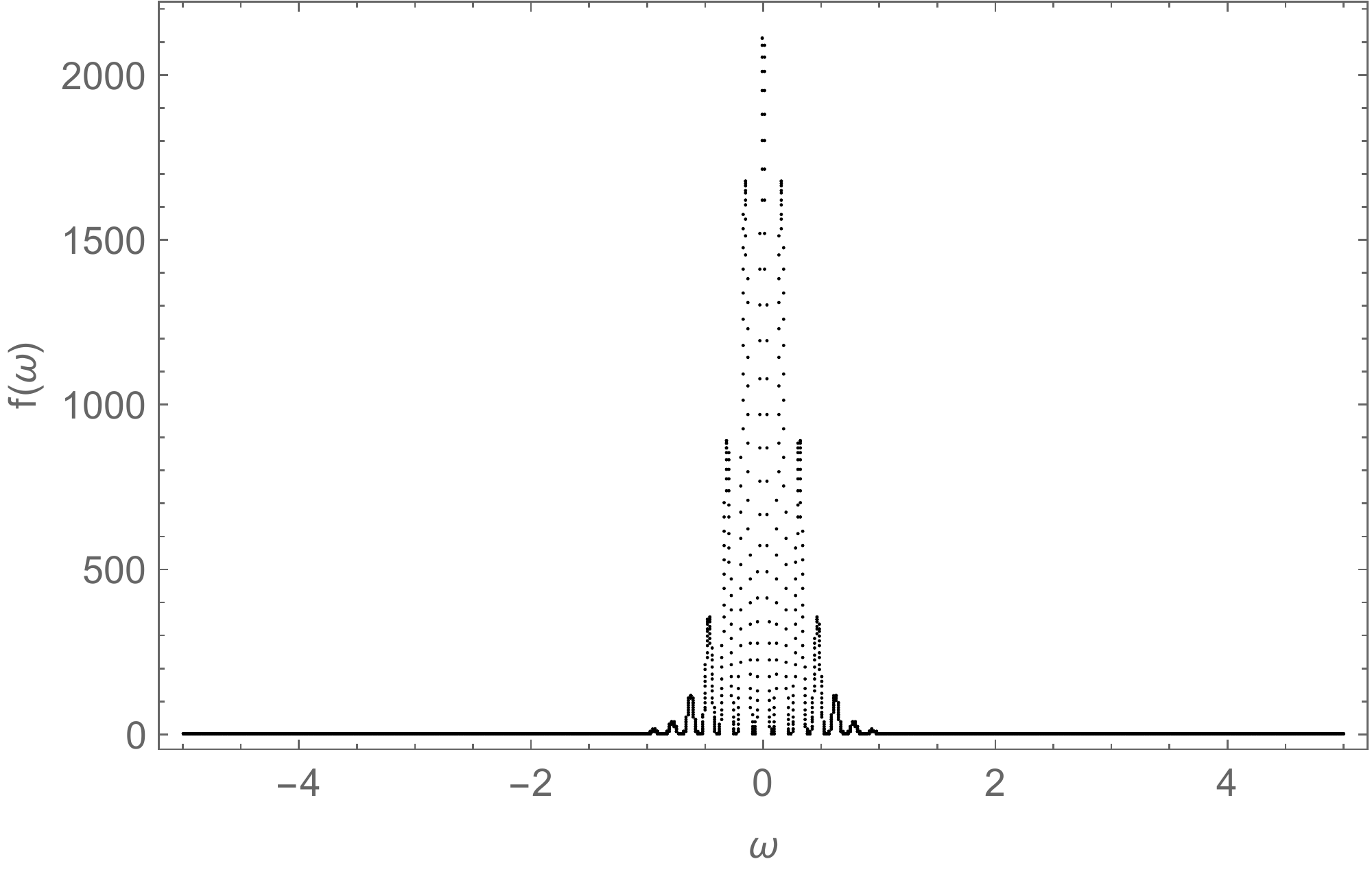}
\quad\quad
\caption{\footnotesize Modal fraction (IR regime) of the dynamical AdS/QCD holographic model, with warp factor compatible with linear Regge trajectories (\ref{cosmologica}), for $S=3$ (left panel) and  $S=6$ (right panel). }
\end{center}
\end{figure}
\begin{figure}[H]
\begin{center}
\includegraphics[width=2.5in]{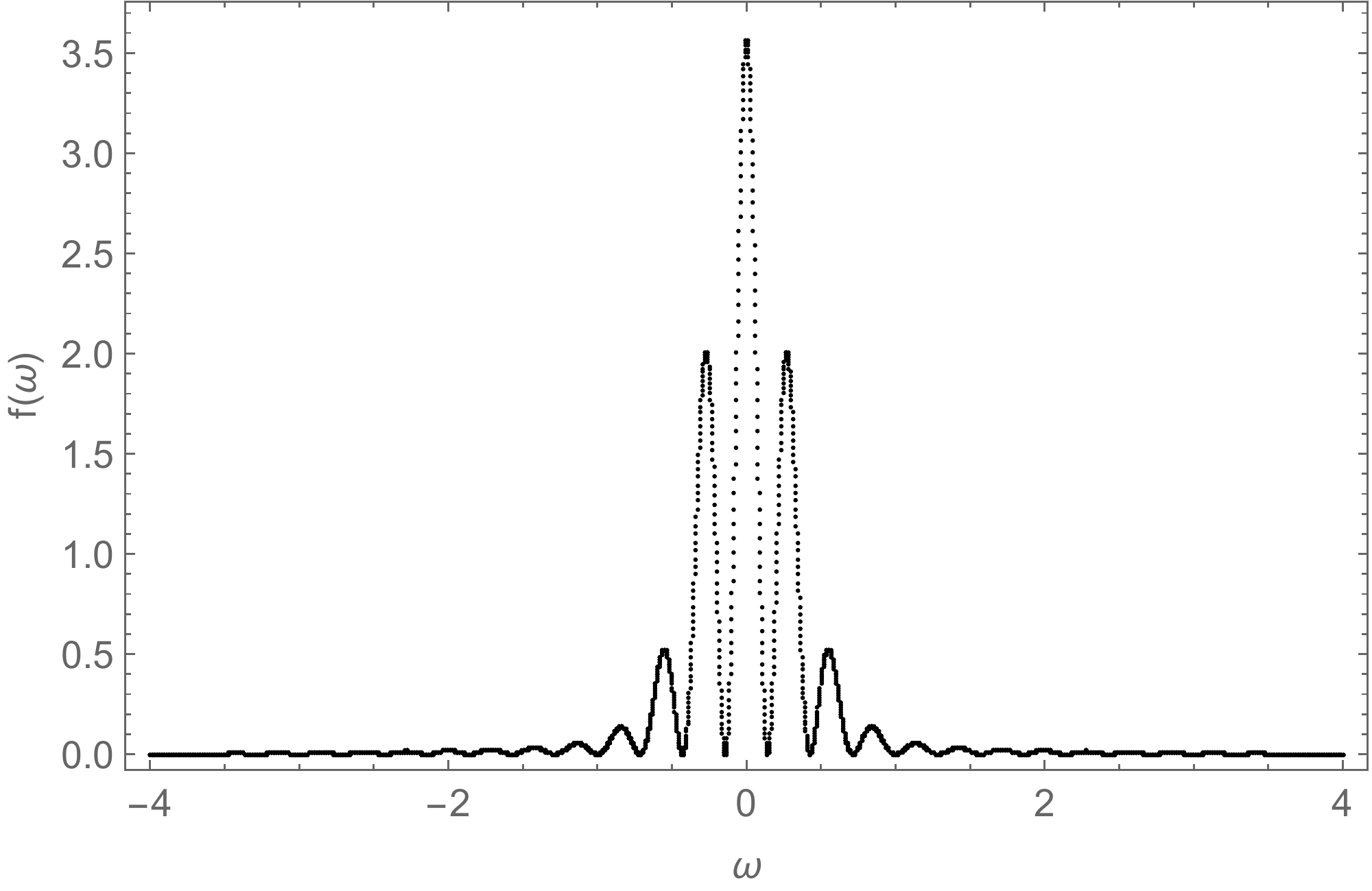}
\includegraphics[width=2.6in]{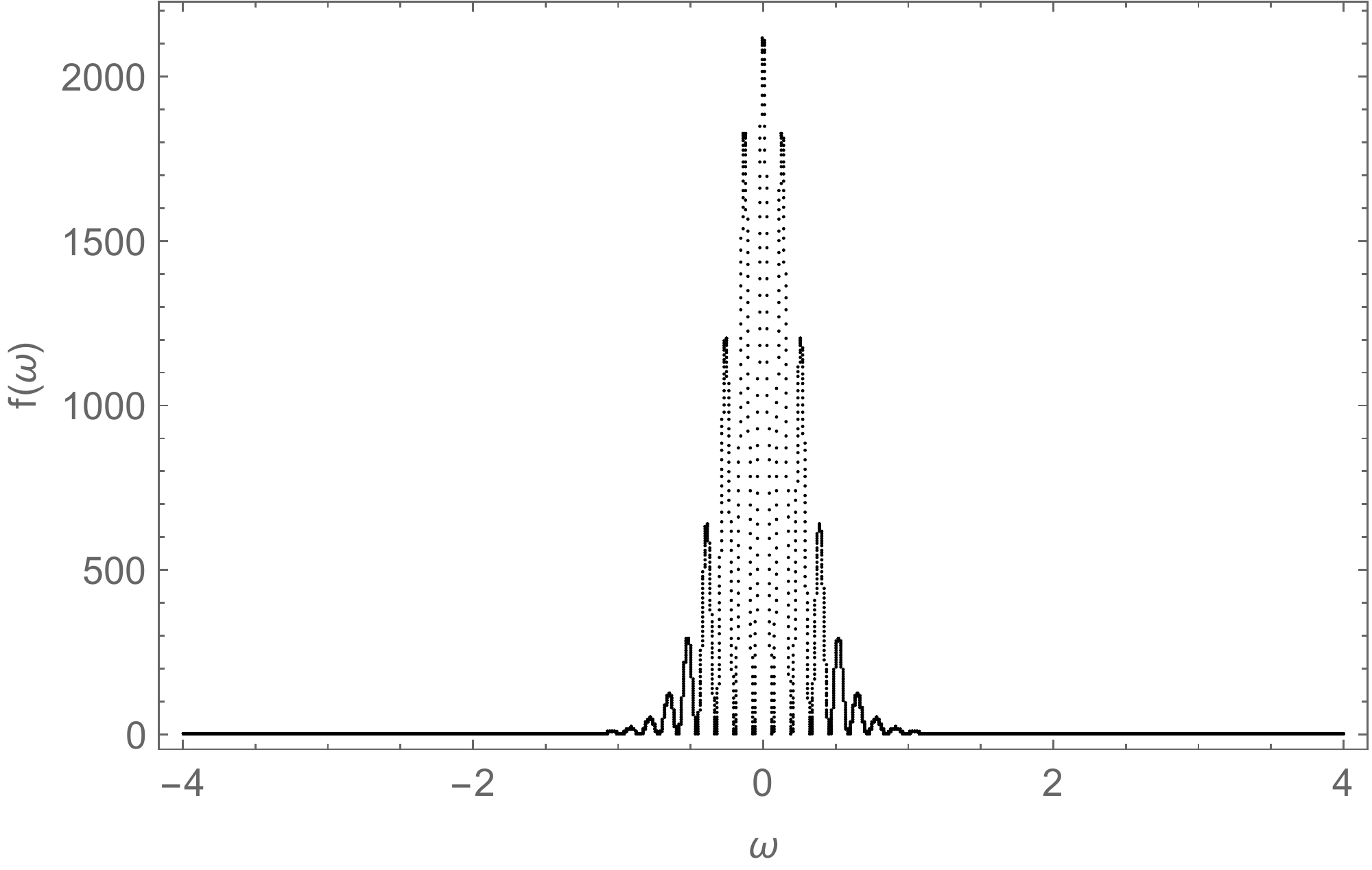}
\quad\quad
\caption{\footnotesize  Modal fraction (UV regime) of the dynamical AdS/QCD holographic model, with warp factor compatible with linear Regge trajectories (\ref{cosmologica}), for $S=3$ (left panel) and  $S=6$ (right panel).}
\end{center}
\end{figure}\vspace*{-1cm}
\end{widetext}

\end{document}